\documentclass[aps,twocolumn,groupedaddress,nofootinbib,nobalancelastpage,nobibnotes]{revtex4-1}
\usepackage{graphicx,color}

\usepackage{amssymb}               
\usepackage{graphicx}
\usepackage{hyperref}
\hypersetup{colorlinks=true, citecolor=red, linkcolor=blue}
\graphicspath{{figures/}}
\newcommand{\caga}[1]{{\leavevmode\color{black}#1}}
\newcommand{\omitt}[1]{{}}
\newcommand{\cambia}[1]{{\leavevmode\color{black}#1}}
\newcommand{\cambiab}[1]{{\leavevmode\color{black}#1}}
\newcommand{\andrea}[1]{{\leavevmode\color{black}#1}}
\newcommand{\andreab}[1]{{\leavevmode\color{black}#1}}
\newcommand{\andreac}[1]{{\leavevmode\color{black}#1}}
\newcommand{\andread}[1]{{\leavevmode\color{black}#1}}

\newcommand{\corr}[1]{{\leavevmode\color{black}#1}}
\newcommand{\att}[1]{{\leavevmode\color{black}#1}}

\begin{document}

\title{Can BL Lac emission explain the neutrinos above 0.2 PeV?}
\author{Andrea Palladino}
\affiliation{Gran Sasso Science Institute, L'Aquila, Italy}
\email{andrea.palladino@gssi.infn.it}
\author{Francesco Vissani}
\affiliation{Laboratori Nazionali del Gran Sasso, Assergi (L'Aquila), Italy, \\ Gran Sasso Science Institute, L'Aquila, Italy}
\email{francesco.vissani@lngs.infn.it}


\begin{abstract}
Multi-messenger astronomy can help to investigate the sources of the high-energy neutrinos observed by the high-energy neutrino telescope IceCube. We consider the hypothesis that the highest energy neutrinos are produced by BL Lacs, arguing that this is not contradicted severely by any known fact. We check the BL Lac hypothesis by searching for correlations between the through-going muon events of IceCube and the BL Lacs of the second catalog of Fermi-LAT (2FHL). We expect $10.2 \pm 2.4$ correlated events but we find that just 1 event has a BL Lac as counterpart. We also assess the probability of observing one multiplet from the same source, finding that the present null result is not yet of critical significance. We conclude that the hypothesis that the BL Lacs are the main emitters of the highest-energy neutrinos observed by IceCube is disfavored at 3.7$\sigma$. We discuss implications and possible ways out; for example, this could work if the angular resolution was $4^\circ$, which is much more than expected.
\end{abstract}

\maketitle

\tableofcontents


\section{Introduction}

\att{The observations of the high-energy neutrino telescope IceCube (we refer to \cite{icecube1,muoni6} for recent results and references therein)
are at least as exciting as those concerning solar  and supernova neutrinos that begun neutrino astronomy and were  recognized by the Nobel prize in physics in 2002. 
The high-energy neutrinos might well eventually constitute a new and rich chapter of neutrino astronomy. 
This precise field of research is still gathering speed, but some relevant elements have already been obtained.}
The highest-energy events that have been observed cannot be entirely attributed to the secondary particles produced by cosmic-ray interactions with the atmosphere. 
These observations can instead be explained invoking the existence of a population of neutrinos of cosmic origin. Interestingly, the arrival directions of the events of highest energy are compatible with an isotropic origin, which suggests that most of them are extragalactic. 
\att{It is now time to form hypotheses on what the sources of these particles could be.

To date, we do not have a clear idea and/or 
astrophysical picture of what the sources  of these, presumably extragalactic, neutrinos are.
Even if one might imagine exotic processes or mechanisms of production,  it seems plausible {\em a priori} that the  IceCube neutrinos originate in some astrophysical environment, rich with cosmic rays and target particles, that allows collisions and therefore the  production of secondary particles. 
In our view, this hypothesis should be investigated thoroughly and the present study can be regarded as a contribution to this discussion.

In order to proceed,  it is natural to rely on synergies with other astronomies, that is, to exploit the potential of a 
``multi-messenger'' investigation.
In fact, the neutrinos from cosmic ray collisions are accompanied  by $\gamma$-rays; therefore, the hypothetical sources of neutrinos must also be sources of $\gamma$-rays. 
The possibility of observing the $\gamma$-rays is by no means guaranteed. For example,  
the  $\gamma$-rays could have a different distribution from those of the neutrinos (e.g., if the neutrinos are emitted 
 isotropically near the nuclear region of an active galactic nucleus (AGN), whereas the $\gamma$ rays emerge as a collimated beam of one jet 
 instead) or could, possibly, 
 be absorbed and/or strongly reprocessed at the source. However the intergalactic medium does not absorb $\gamma$-rays with  energies below 100 GeV or so. 
Thus, through the use of results from astrophysical
modeling, there is hope for revealing a correlation between these $\gamma$-rays and the observed neutrinos.}
 
 In this connection, Fermi-LAT, that has obtained a relatively complete survey of the $\gamma$-ray sky below a few 100 GeV, 
 has produced results that are of special interest. 
In the region $E_\gamma>10$ GeV, the brightest objects are the \andread{blazars (we refer to e.g., \cite{p} for a review), radio-loud AGN whose relativistic jet is 
closely aligned with the line of sight. They are broadly classified 
as BL Lacs (namely BL Lacertae) and as flat spectrum radio quasars (FSRQ).  
BL Lacs are characterized  by featureless optical spectra (i.e., lacking strong emission/absorption lines).
A finer classification of BL Lacs is considered below.}

Blazars 
are widely considered to be  promising candidates for high-energy neutrino emission (\cite{Protheroe:1996uu,Atoyan:2001ey,beacom,murase3,halzen,
murase2,murase1,Resconi1,Resconi2,
messicani,icecube1,icecube2,winter}). \att{Moreover,
certain analyses of IceCube data (we refer to  \cite{Resconi2} for a very recent and comprehensive work) suggest that a 
fraction of the events seen by IceCube could be attributed to blazars.}

\att{In recent works (\cite{tavecchio14b,Resconi3,tavecchio14a,TavecchioRighi})  focus was put 
on the subclass of blazars called BL Lacs; the present study also concentrates on this hypothesis.
In the models of Tavecchio et al., (we refer to in particular \cite{tavecchio14b,tavecchio14a,TavecchioRighi}),  
the secondary high-energy particles (neutrinos and $\gamma$-rays) are to a good extent produced in alignment with the direction of the jet   (spine-layer model) and therefore one expects a tight correlation of the 
signals. This highly definite astrophysical setup adds motivations to the present investigation.}

\begin{figure}[t]
\begin{center}
\centerline{\includegraphics[width=8cm]{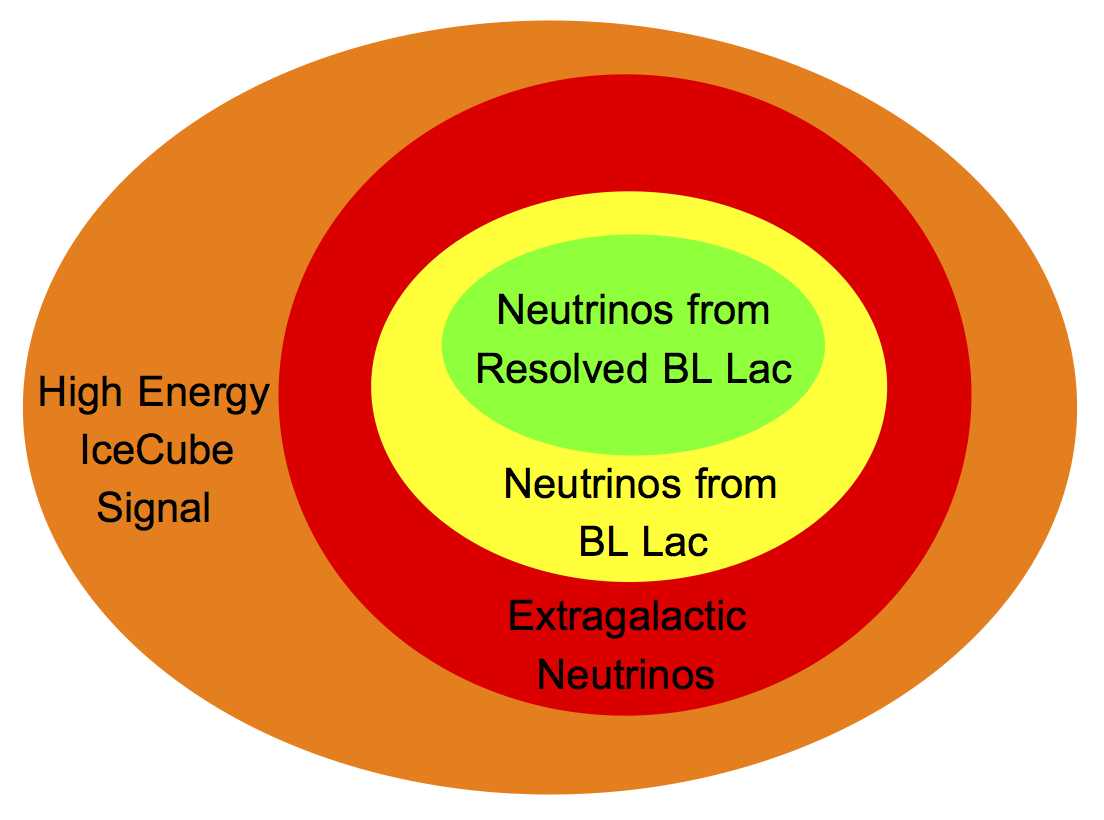}}
\caption{\small\it Scheme of the present investigation: 
We assume that {\em a part of the high energy neutrino signal}
seen by IceCube includes extragalactic neutrinos. We investigate whether the extragalactic emission 
can be attributed largely or fully to BL Lac emission, relying on the fact that 
a large fraction of BL Lac\corr{s}, as discussed in Sect.~\ref{pq}, is observed by Fermi-LAT.}
\label{figa}
\end{center}
\end{figure}

At this point we can formulate the question that we examine and discuss: 
\begin{quote}
 {\em Are the highest energy 
extragalactic neutrinos seen by IceCube fully attributable to BL Lacs?}
\end{quote}
This hypothesis, presented in a slightly more general form in 
Fig.~\ref{figa}, is investigated in this work.
We note that neutrinos and the $\gamma$-rays are observed at rather different energies \att{in IceCube 
and Fermi-LAT, respectively,}
and 
that the details of the connections greatly depend upon on uncertain theoretical modeling. Thus, 
the investigation is bound to proceed on  phenomenological grounds.
We refer to \cite{icecube1} for a similar and complementary study of blazars and neutrinos.

The outline of this work is as follows. 
In Sect.~\ref{pq} we define the context of the discussion; we discuss under which conditions   
the above hypothesis is viable. 
Then we perform the crucial test,  
by searching for correlations between extragalactic neutrinos and  the subclass of
high-energy emitting  BL Lacs that are identified astronomically. We argue in Sect.~\ref{spacorr} that the number of observed correlations is too small and that the hypothesis  is not supported.
We also derive the  expectation on multiplets of neutrino events and compare this with the observations   (Sect.~\ref{multi}). 
\corr{We discuss, in Sect.~\ref{conc}, the results and the possible ways out from the conclusion that the observed neutrinos with energy above 200 TeV receive only a minor contribution from BL Lacs.  }

\section{Extragalactic neutrinos and BL Lacs}
\label{pq}

Before examining the connection, we need to discuss a few important questions:
1)~Which set of neutrino data gives us the best chance of extracting and identifying the  extragalactic part of the signal;
2)~What is the fraction of the extragalactic neutrinos of IceCube \andreac{that} should correlate with BL Lacs;   
3)~How do we compare with the limit on neutrino emission from blazars obtained by IceCube in \cite{icecube1}?

\subsection{Importance of the high-energy through-going events}
\att{The largest high-energy neutrino telescope today in operation is IceCube.  This observatory consists in a 
volume of about 1~km$^3$ of Antarctic ice, monitored by strings of phototubes. When the neutrinos 
interact inside or nearby the detector, they produce charged particles that allow for the detection of the presence of 
high-energy neutrinos 
and the measurement, to a certain extent, of their energy, direction, and flavor (i.e., electron, muon or tau). There are two main classes of signals detected in 
IceCube: 1)~The {\em through-going events}, that are muons (or antimuons) due to neutrinos interacting outside the detector; and 2)~the {\em high-energy starting events} (HESE), whose vertex is instead contained inside the instrumented volume. Among HESE, one distinguishes those events where there is a track, associated to charged-current 
muon neutrino interactions,
and those where the track is absent and the energy is deposited in a small region, that are associated to the other 
types of flavor and interactions.}

Recently IceCube has published a \att{relatively large} set of 29 through-going events, 
collected over six years (\cite{muoni6}) and with very high energy; all
\andreac{of them} have energy $\ge 0.2$ PeV and 
the highest energy event seen to date
\att{corresponds to a neutrino 
with more than 4.5 PeV}  and 
belongs to only this \andreac{dataset}. 
These  through-going events provide strong support for the observation of a signal of cosmic neutrinos 
previously claimed by IceCube with the 
HESE dataset. When comparison is possible, 
the two datasets are consistent with a common, simple interpretation.
In this paragraph we discuss the three main reasons  
why  we consider that  this specific dataset is particularly important 
for the investigation of the BL Lac hypothesis. 

  \paragraph*{Energy distribution}
  The neutrino signal  corresponding to the  through-going event  dataset 
  is compatible 
  with a single power-law distribution with a spectral index
  $\gamma=2.13\pm 0.13$ (\cite{muoni6}). 
   It is remarkable that this distribution 
  agrees well with the high-energy part of the HESE dataset, as argued in
  \cite{gal2} and eventually proved in \cite{muoni6}.\footnote{A featureless and hard power-law spectrum would suggest a $pp$ production mechanism   for neutrino production, 
  rather than the usually expected $p\gamma$ mechanism. However, it is not possible to say that the $p\gamma$ mechanism is unavoidable for BL Lac\corr{s}. 
 Moreover, the spectrum is measured only in a small range of energy between 200 TeV and some PeV. 
 The high energy part of the spectrum is not yet sensitive to  mechanism of production, and this is not a critical aspect of the BL Lac hypothesis.
  We refer to \cite{winterpg1,winterpg2} for a detailed discussion of the photohadronic interaction.}
  The bulk of the HESE data, that includes the events 
  collected at lower energies, indicates a different spectral index  instead, closer to $\gamma= 2.5$.
  This feature of the HESE data can be attributed  to the onset of another component of the signal with  
  lower energy that is plausibly not entirely of extragalactic origin, as discussed in  \cite{s,gal1,gal2}.  
  We note that a relatively hard distribution of extragalactic neutrinos can be extrapolated at lower energies without particular difficulty, 
  whereas, if the extragalactic neutrinos were distributed  as $E^{-2.5}_\nu$ at low energies, this would imply 
  an excessive amount of $\gamma$-rays (\cite{armanvis}), or too many low-energy track events, 
  similar to those expected from prompt atmospheric neutrinos and  for which we have no evidence (\cite{gal2}). 
    Thus, in order to remain cautious for 
  what concerns the interpretation of the extragalactic neutrino distribution
  (i.e., the population of neutrinos that is relevant to assess the BL Lac\corr{s} contribution)
  we deem that it is preferable to focus the discussion on the high-energy part of the spectrum.

   \paragraph*{Northern hemisphere} 
   The through-going events originate from muon neutrinos/antineutrinos that are converted 
   into muons/antimuons near the detector. 
   For this class of signal events, the Earth works, at once, as a neutrino converter and as a screen for the atmospheric muons. Thus, most of the events come from below the horizon; in the case of IceCube this means the Northern hemisphere. In this hemisphere, it is expected that the  contribution of the signal due to galactic sources 
   is small or absent. This statement is even more true for the specific dataset used in \cite{muoni6}, which satisfies also the 
   high-energy criterion discussed immediately above. We refer to \cite{s,gal1,gal2} for further discussion.

\begin{figure}[t]
\begin{center}
\centerline{\includegraphics[width=8cm]{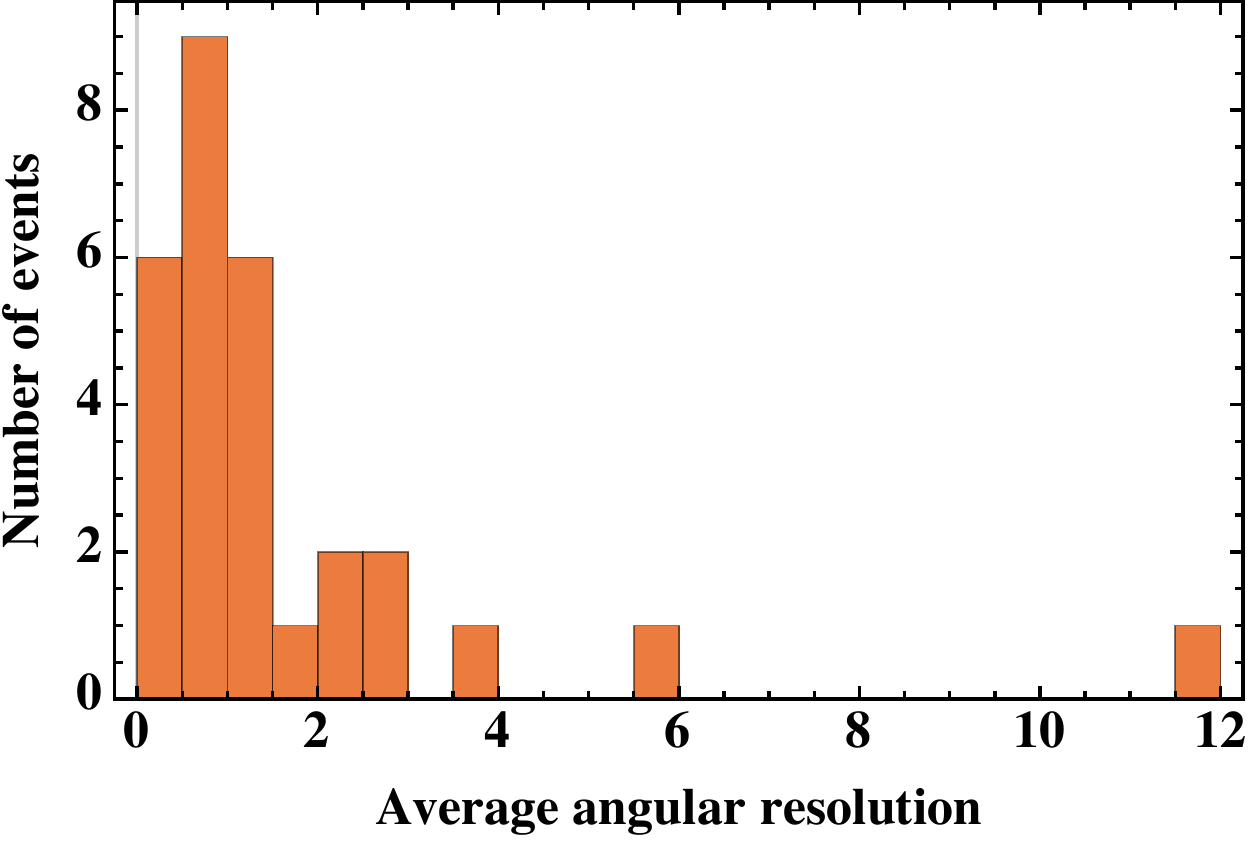}}
 \caption{\textit{Values for the 29 through-going events of the average angular resolution, defined as described in Eq.~\ref{defello}. We note the two outliers, discussed in the text. }}
\label{figb}
\end{center}
\end{figure}

   \paragraph*{Pointing of through-going events} 
    The last and most important characteristic of through-going events is that their arrival directions are known relatively well.
    This is a general feature of events of the track type, although in some cases the precision is bad, similar to the one of 
 the HESE, shower events. In Table 4 of \cite{muoni6}, the declinations $\delta$ and the right ascension $\alpha$ are given along with their 
 asymmetric errors $\delta_{\pm}$ and $\alpha_{\pm}$, at the 
 50\% CL and at 90\% CL. We  have checked that the errors in the two 
 cases scale roughly according to a Gaussian distribution   but we 
 use, as a rule, the more conservative case (90\% CL) in our analysis.
 The angular span of these two angular coordinates is given by the sum of the upper and lower ranges, that are, in general, different: 
 $\Delta\delta(i)=\Delta\delta_+(i) + \Delta\delta_-(i)$ and likewise  $\Delta\alpha(i)=\Delta\alpha_+(i) + \Delta\alpha_-(i)$ where $i=1,2,3....29$. 
 Each event subtends a solid angle of 
 $\Omega(i)=\Delta\delta(i) \times \Delta\alpha(i)$, that, according to the quoted confidence levels, 
 is expected to include the true direction with a confidence level of $0.9^2=81$\%.
 We note that the total angular size spanned by the neutrino events is small 
\begin{equation}
\label{caliban}
\sum_{i=1}^{29} \Omega(i)=3.5\%\times (2 \pi)
,\end{equation}
where we compare with the solid angle subtended by Northern sky.
 It is also convenient to define a single (linear) angle 
 that quantifies the region around the individual event as follows,  
 \begin{equation} \label{defello}
\theta(i) \equiv \sqrt{\frac{\Omega(i)}{\pi}} 
.\end{equation}
We refer to this angle as the ``average angular resolution''.
A histogram of values based on the 29 events 
is shown in Fig.~\ref{figb}. We note the presence of two outlying events, whose direction is identified only poorly.
These are event number 6 and event number 14 in Table~4 of \cite{muoni6}.
In order to be more quantitative, we note that 1)~the event number 6 alone covers $61\%$ of the solid angle spanned by neutrinos
given in Eq.~\ref{caliban}; and that
2)~if we exclude the two outliers, and consider the angle spanned by the remaining 27 events, 
$3.5\%\to 0.9\%$.
\andreac{Considering the angular resolutions shown in Fig.\ref{figb}  and excluding the two outliers, we have the average angular resolution  $\bar{\theta}=1.2^\circ \pm 0.9^\circ$.} 

       \paragraph*{Investigation of the origin of the extragalactic component}
         The first reason why the 29 through-going events, the largest sample of high-energy events 
  collected by IceCube, are so important to us is that they are compatible with an extragalactic origin.  
 The fact that these events come from the Northern hemisphere  gives  us further confidence supporting the hypothesis that 
    the cosmic neutrino signal that contributes to this dataset     has an extragalactic origin.  
    The latter feature (i.e., pointing) offers us the chance to use these data to identify the sources of the events; we 
    use it in Sect.~\ref{spacorr} to test our assumption on the high-energy neutrino emission 
    from BL Lac.

    \subsection{Fraction of correlated neutrino events}
    
    \paragraph*{Modeling of BL Lac}
    The $\gamma$-rays telescopes and 
in particular Fermi-LAT can see individually 
only a fraction of the BL Lac\corr{s}. The rest corresponds to objects that are too dim (i.e., too far and/or too under-luminous) 
to be observed; we refers to these as un-resolved BL Lacs.
This fraction is of the order of one and it is widely believed  that 
the class of BL Lacs  that are 
un-resolved gives an important contribution to the ``diffuse'' $\gamma$ ray emission at the highest energies. 

It is possible to be more precise thanks to the model of  \cite{ajello}. 
\att{In this paper, a detailed parameterization of the 
luminosity function of the BL Lac\corr{s} distribution, 
differential in $0.1-100$ GeV luminosity ($L_\gamma$), redshift ($z$) and photon index ($\Gamma$),
\begin{equation}
\frac{\partial^3 N}{\partial L_\gamma \partial z \partial \Gamma}
,\end{equation}
is studied. The parameters are obtained, within errors, 
thanks to the observational data. The result depends on modeling and in the following we  adopt the result from the 
two best models  of \cite{ajello}, there referred to as LDDE (luminosity-dependent density evolution) 
and PLE (pure luminosity evolution). 

Exploiting the results of this analysis it is 
not difficult to calculate the total flux of $\gamma$-rays due to BL Lacs, 
\begin{equation}
\phi_{\mbox{\tiny tot}}^\gamma = 
\int dL_\gamma dz d\Gamma\ \frac{\partial^3 N}{\partial L_\gamma \partial z \partial \Gamma}\
\frac{dF}{dE_\gamma}(L_\gamma,z,\Gamma;E_\gamma) 
\label{macacon}
,\end{equation}
where $dF/dE$ is the differential 
flux of the single power-law source, that integrated over the energies gives,
\begin{equation}
F(L_\gamma,z,\Gamma)=\frac{\mathcal{N}_\gamma}{4\pi D_c^2(z) } (1+z)^{-\Gamma}
\mbox{ with } \mathcal{N}_\gamma=\frac{L_\gamma}{\langle E_\gamma\rangle} 
,\end{equation}
where $D_c(z)$ is the comoving distance and the factor $(1+z)^{-\Gamma}$ accounts for the redshift.
For a sample plot of the total flux  in the case of LDDE model,  
obtained using Eq.~\ref{macacon}, see 
Fig.~3 of \cite{armanvis}.

The result of the calculations of the quantities relevant for the present papers,  
namely the $\gamma$-ray fluxes from resolved and un-resolved BL Lacs, 
are given in \cite{armanvis}, which is fully based on the work of \cite{ajello}.
The flux of $\gamma$-rays due to the resolved BL Lacs is, 
$\phi_{\mbox{\tiny res}}^\gamma=8.5\times 10^{-7}$ ph/(cm$^2$ s sr),
with a small uncertainty.
The flux of the un-resolved BL Lacs 
is given by 
$\phi_{\mbox{\tiny unres}}^\gamma=\phi_{\mbox{\tiny tot}}^\gamma-\phi_{\mbox{\tiny res}}^\gamma$,
thus, its value depends on theoretical modeling. 
Table~2 of \cite{armanvis} gives the result for the  two models mentioned above:
$\phi_{\mbox{\tiny unres}}^\gamma(\mbox{LDDE})=8^{+2}_{-1.3}\times 10^{-7}$ ph/(cm$^2$ s sr)
and 
$\phi_{\mbox{\tiny unres}}^\gamma(\mbox{PLE})=10^{+2.1}_{-1.7}\times 10^{-7}$ ph/(cm$^2$ s sr).
We use conservatively 
the upper range of the largest prediction (PLE),  
$\phi_{\mbox{\tiny tot}}^\gamma<20.6\times 10^{-7}$ ph/(cm$^2$ s sr) 
and the lower range of the smallest prediction  (LDDE), 
$\phi_{\mbox{\tiny tot}}^\gamma>15.2\times 10^{-7}$ ph/(cm$^2$ s sr),
thereby finding the fraction of the $\gamma$ ray flux  due to resolved BL Lacs,
\begin{equation}
f_\gamma=\frac{\phi_{\mbox{\tiny res}}^\gamma}{\phi_{\mbox{\tiny tot}}^\gamma}= 0.5 \pm 0.1
\label{fracbl}
.\end{equation}
Therefore, this fraction is relatively well-known, 
the uncertainty being about 20\% 
(we note incidentally that the  un-resolved flux 
can account for a large part of the  diffuse flux observed above 10 GeV as argued in \cite{ajello}). }

An important remark is as follows. 
On the one hand, we would like to have information on $\gamma$ ray emission at the highest energy, but on the other hand, 
the Universe becomes  opaque  to $\gamma$ rays above 100 GeV, and this limits the useful observational window. 
From this point of view, the subset of BL Lacs characterized by an intense emission above 50 GeV, collected in the catalog 2FHL,  
\cite{2fhl}, is particularly interesting for us and we use it in the following.  
A complete model of the cosmological evolution of these objects is not yet available; however, the 
BL Lacs of  the 2FHL catalog belong mostly to the subclass of the high-frequency synchrotron peak (HSP) that 
were studied in \cite{ajello}. We note that the HSP have a fast cosmological evolution, that is, they are more abundant at low redshift values
(we refer in particular to Fig.~11 of  \cite{ajello}). From this consideration, it is expected  
that the visible fraction of the BL Lac\corr{s} included in \cite{2fhl} is possibly even higher than 0.5. This conclusion is very important in view of the subsequent analysis.

\paragraph*{Hypothesis on  neutrino emission}
In most theoretical models, the neutrino luminosity is proportional to the
$\gamma$ ray luminosity; we refer to  
\cite{stecker}; 
\cite{Protheroe:1996uu,Atoyan:2001ey,beacom,halzen,murase3,murase2,murase1,TavecchioRighi}.  Therefore, we assume that the 
fraction of the extragalactic neutrino flux $f_\nu$ that corresponds to resolved BL Lac\corr{s} (i.e., the tagged fraction) equates the fraction determined immediately above\footnote{\andread{We note that we do not need to introduce a proportionality coefficient between the flux of $\gamma$-rays and the flux of neutrinos, since we are only interested in the fraction of visible flux with respect to the theoretical (expected) total flux. E.g., in \cite{TavecchioRighi} the almost constant coefficient $\phi_\nu=0.46 \phi_\gamma$ is obtained in a BL Lac scenario, but it disappears from the ratio of Eq.~(\ref{fracbl}). }},
\begin{equation}
f_\nu=f_\gamma
,\end{equation}
and we assume that this is true in particular for what concerns the 
BL Lacs that emit the very-high-energy events observed by means of 
through-going muons.

Since we use the integrated quantities, this inference should be relatively stable and not crucially dependent upon the detailed connection between neutrinos and gamma rays.   
We argued that the fraction of neutrinos due to the BL Lac\corr{s} 
that emits $\gamma$-rays at the highest observed energies 
(in particular those included in 2FHL catalog)
is not underestimated by $f_\gamma$. In other words, when we use Eq.~\ref{fracbl} along with the 
 high-energy neutrino signal, we obtain the 
  minimum number of expected correlations 
 with the BL Lac\corr{s} in the 2FHL catalog.

    \subsection{Comparison with the bound on neutrinos obtained by IceCube}
    
      \att{Recently IceCube has published a study (\cite{icecube1}) concerning neutrinos from blazars where a limit 
     on the fraction of the signal neutrino events observed and attributable to these astrophysical objects was obtained.
    This study is relatively similar in spirit to the present work, namely, it is a search for correlations between certain classes of astronomical objects and the observed neutrinos, 
   hypothesized to originate from these objects.
   Thus, it is especially important} to highlight the differences between our approaches and theirs:
   \begin{enumerate}
\item IceCube collaboration focusses on the study of blazars whereas we prefer to be more specific and emphasize the comparison with the BL Lacs instead. 
\item The result of IceCube is based on data of energy above 10 TeV, \att{that is, well below the lowest 
energy of the through-going muon dataset that we use, 200 TeV.}
\item Finally and most importantly, we consider the total set of BL Lacs, not only those that are resolved, since it is presumable that also the ones that are not resolved emit high energy neutrinos. 
\end{enumerate}
 Therefore, we examine here the bound of IceCube 
    from a different perspective.  


\begin{figure*}[t]
\centering
\includegraphics[width=0.9\textwidth,angle=0]{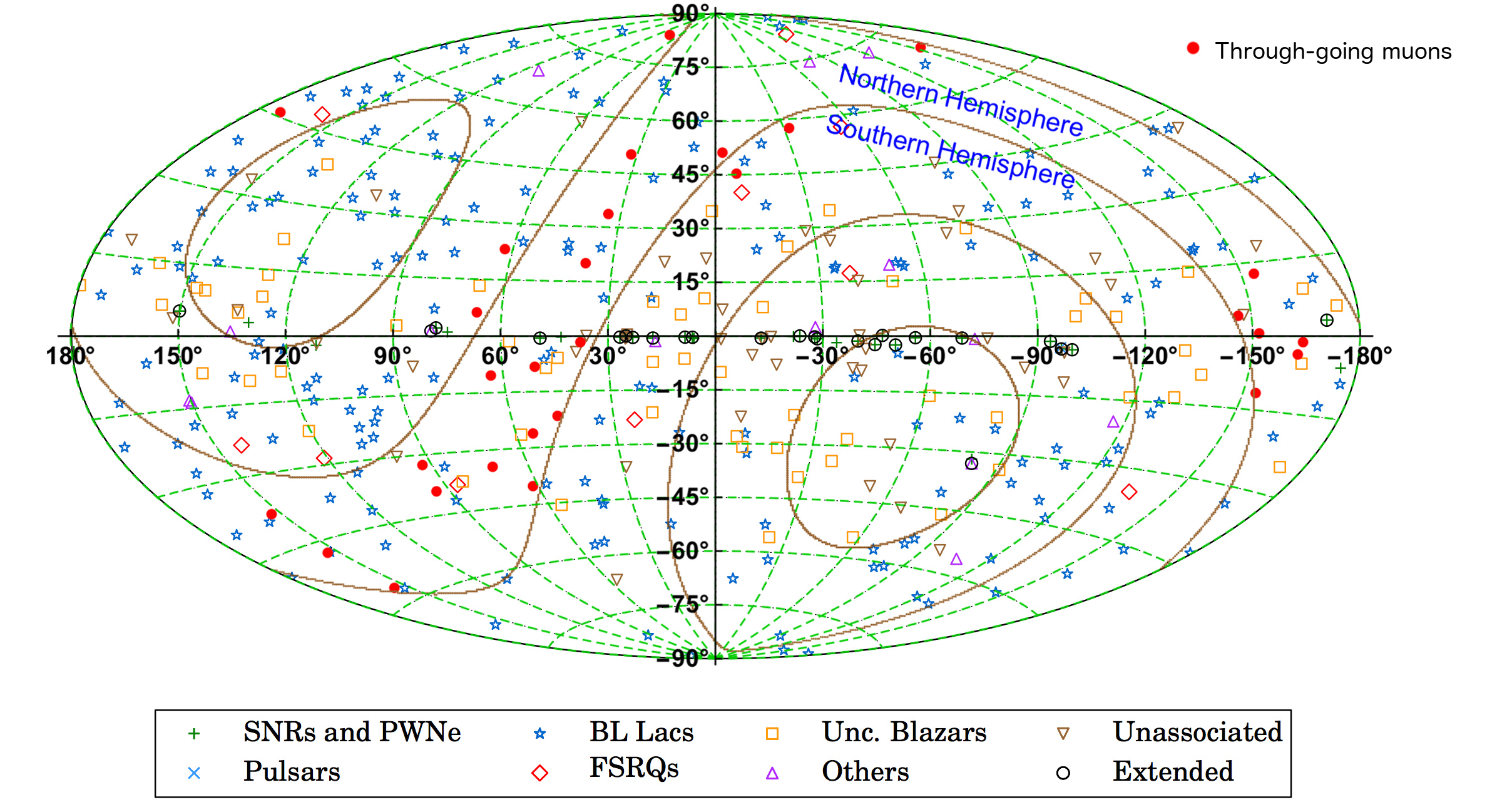}
\caption{\textit{Map, in Galactic coordinates, of the $\gamma$-ray sky observed by Fermi-LAT, from \cite{2fhl}. {The red points indicate} the 29 through-going muons, with a deposited energy above 200 TeV, observed by IceCube over 6 years; {their angular uncertainties are not represented in this figure, but are taken into account in the calculations}. The brown lines represent intervals of declination of $30^\circ$.  }}
\label{map}
\end{figure*}


We consider the simplest description of the 
neutrino signal that explains  the through-going muons flux. This is a power law with  $\alpha=2$, a value admitted within  
1$\sigma$, as illustrated in Fig.~6 of \cite{muoni6}. From the same figure we find that the normalization of 
flux of $\nu_\mu$ and $\bar{\nu}_\mu$ is equal to
\begin{equation}
\phi^{\nu_\mu+\bar{\nu}_\mu}
_{\mbox{\tiny through-going}}
= (6 \pm 2)  \left(\frac{E}{\rm 1 \ GeV} \right)^{-2} 
\frac{\rm 10^{-9}\ GeV}{\rm cm^2 \ s \ sr}
,\end{equation}
with an uncertainty of about 30\% on the normalization. 
This describes the extragalactic neutrino flux that we 
assume to be fully explained by BL Lac emission. In order to calculate the fraction of the neutrino flux due to 
resolved BL Lac\corr{s}, we simply have to multiply it by the fraction given in  
Eq.~\ref{fracbl}, obtaining
\begin{equation}
\phi^{\nu_\mu+\bar{\nu}_\mu}_{\mbox{\tiny res}}\mbox{(theo.)}= (3 \pm 1.2) \left(\frac{E}{\rm 1 \ GeV} \right)^{-2} \frac{\rm  10^{-9}\ GeV}{\rm cm^2 \ s \ sr}
,\end{equation}
where we summed the uncertainties in quadrature. 
The bound obtained by IceCube is reported in Table~3 of \cite{icecube1} for the case where $\alpha=2$,  ``All 2LAC Blazars'' and ``equal weighting''. 
The bound is 
\begin{equation}
\phi_{\mbox{\tiny res}}^{\nu_\mu+\bar{\nu}_\mu}(\mbox{obs.})< (4.7 \pm 0.5) \left(\frac{E}{\rm 1 \ GeV} \right)^{-2} \frac{\rm  10^{-9}\ GeV}{\rm cm^2 \ s \ sr}
,\end{equation} 
where we use the 1$\sigma$ uncertainty instead of the 90\% confidence level (C.L) quoted in the table. This bound is not incompatible with the 
hypothesis that BL Lacs produce all the extragalactic neutrino flux.
We note that the second reason 
why we have chosen to discuss the spectral index $\alpha=2$ for the through-going muons flux is that this value permits a direct comparison between the measured fluxes and the bound. The bound of IceCube becomes tighter by increasing $\alpha$
and for $\alpha=2.2$ excludes that the blazars of \cite{2fhl}  emit more 
that 50\% of the neutrinos (\cite{icecube1}). However, in view of Eq.~\ref{fracbl}, 
this is not incompatible with the hypothesis that the BL Lacs
(those resolved and unresolved)  
account for the full extragalactic neutrino emission. 
    \att{Similar analyses, using a wider set of catalogs for the 
$\gamma$-ray sources, are presented 
in \cite{Resconi2}; the results are slightly tighter but comparable with those of IceCube.}

\section{Spatial connection between BL Lacs and high-energy neutrinos}
\label{spacorr}
Here we investigate whether the counterparts of high-energy neutrino events contained in \cite{muoni6} are the BL Lacs. 
\label{corrsec}

\subsection{The number of expected correlations}
\label{liksig}
\att{Cosmic rays that hit the terrestrial atmosphere yield secondary particles that act as  background events, that is, they contaminate the  
observational sample: These are (1) muons that come mostly from 
the sky above the detector and affect the  interpretation of the HESE dataset but not much the through-going muons  and 
(2) neutrinos and antineutrinos that have a different (softer) energy spectrum than the signal 
and are mostly muon neutrinos and antineutrinos at relatively low energies. The role
of neutrinos in the decay of charmed mesons is not known precisely and this creates uncertanty in the inferences at a few 100 TeV; however it is possible to use the data themselves to obtain a bound 
on this component that is not far from the current theoretical predictions as shown in \cite{muoni6}.

In order to calculate the number of expected correlations,}
the first step is the evaluation of 
the number of tracks that can be attributed to the signal, that is, to astrophysical neutrinos. This can be done using 
Table~4 of \cite{muoni6}, where  the \lq\lq signalness\rq\rq $s_i$
for each event $i=1,2...29$ is given. This quantity 
is defined as {the ratio of the astrophysical expectation over the sum of the atmospheric and astrophysical expectations for a given energy proxy and the best-fit spectrum.}
We use these values of $s_i$ to build a likelihood function that  estimates the total number of events that can be attributed to the signal, or equivalently, the number of events that have to be attributed to the background.
This is obtained expanding the polynomial $\mathcal{P}(x)$,
\begin{equation}
\mathcal{P}(x)=
\prod_{i=1}^{29} [s_i x+(1-s_i)] = \sum_{n=0}^{29} p_n x^n
,\end{equation}
where the coefficient $p_n\le 1$ represents the probability that there are exactly $N_{\mbox{\tiny signal}}=n$ signal events; the consistency condition
$ \sum_{n=0}^{29} p_n=1$ holds true.
In this manner, we found that the through-going event 
dataset contains a number of signal events equal to
\begin{equation}
N_{\mbox{\tiny signal}}=20.4 \pm 2.4
,\end{equation}
where we quote the uncertainty at $1\sigma$. 
When we combine this information with that provided by Eq.~(\ref{fracbl}),
we find that the  
number of neutrino events, which should show a spatial correlation with BL Lacs, is simply
\begin{equation}
N_{\mbox{\tiny corr}}=f_\nu \times N_{\mbox{\tiny signal}}
\label{evcorr}
.\end{equation}
The contribution to its uncertainty is given both by the uncertainty on the experimental data $N_s$ and by the uncertainty on $f$, as follows:
\begin{equation}
\Delta N_{\mbox{\tiny corr}}= N_{\mbox{\tiny corr}} \sqrt{\left(\frac{\Delta N_{\mbox{\tiny signal}}}{N_{\mbox{\tiny signal}}}\right)^2+\left(\frac{\Delta f_\nu}{f_\nu}\right)^2}
.\end{equation} 
This results in
\begin{equation}
N_{\mbox{\tiny corr}}=10.2 \pm 2.4
\label{sigexp}
,\end{equation}
that \andreac{has} an uncertainty of 25\%. Therefore, we assume that the likelihood $\mathcal{L}_{th}(n)$ that gives the expected number of correlations is a Gaussian function, with mean value $\mu=10.2$ and standard deviation $\sigma=2.4$.


\subsection{Number of observed correlations estimated adopting IceCube uncertainties} 
In order to search for the counterparts of the high-energy neutrinos, we use 
the coordinates of the through-going muons and also the uncertainties\footnote{In the table, only statistical uncertainties are reported. We assume here that the systematic uncertainties give a sub-dominant contribution to the total uncertainties and discuss this point in the following.} listed in Table~4 of \cite{muoni6}
along with the 2FHL catalog of the $\gamma$-ray sources 
of the Fermi-LAT collaboration (\cite{2fhl}). These data are shown in 
Fig.~\ref{map}. We note that most of the neutrino events come from the region between $0^\circ \leq \delta \leq 30^\circ$; this is consistent with the fact that \andreac{other neutrinos in this sample} are largely absorbed 
in the Earth, \andreac{since they have high declinations and 
high energies (namely, a reconstructed energy of muons above 200 TeV): \cite{muoni6}.}

When we compare them with the positions of the BL Lacs in \cite{2fhl}, we find that \andreab{there is only one correlation within the 81\% C.L. 
More precisely, \andreab{the neutrino event number 6 of Table~4 of \cite{muoni6} 
can be associated with the BL Lac \textit{J1725.1+1154} or the BL Lac \textit{J1555.7+1111}.}
(As already discussed, we used the confidence level interval reported in Table~4 of \cite{muoni6}. This corresponds to 90\% C.L. \andreac{for declination and right ascension}, and therefore the two \andreac{dimensional} confidence level is slightly less, i.e. $0.9 \times 0.9$ = 81\%.})


It is important to underline that the direction of the neutrino event number 6 is not well reconstructed. In fact this event is one of the two outliers illustrated in Fig.\ref{figb} and its angular uncertainty is of the order of \andreac{$10^\circ$}, much larger than the typical value of about $\sim 1^\circ$. Therefore this correlation could be attributed to the poor angular uncertainty of the events. However, in order to be conservative, \andreab{we do not rule out this correlation  in the rest of this analysis.}


\begin{table}[b]
\caption{\textit{Number of correlations as a function of $k$ and the corresponding confidence level (C.L.)}}
\begin{center}
\begin{tabular}{ccc}
\hline
$k$ & C.L. & N. of correlations \\
\hline
1 & 0.466 & 0\\
1.5 & 0.751 & 1  \\ 
2 & 0.911 & 1 \\
2.5 & 0.975 & 2 \\
3 & 0.995 & 4  \\
3.5 & 0.999 & 6  \\
4 &  $>$ 0.999 & 6 \\
\hline
\end{tabular}
\end{center}
\label{tab:kcl}
\end{table}%

\begin{figure}[t]
\centering
\includegraphics[width=0.45\textwidth,angle=0]{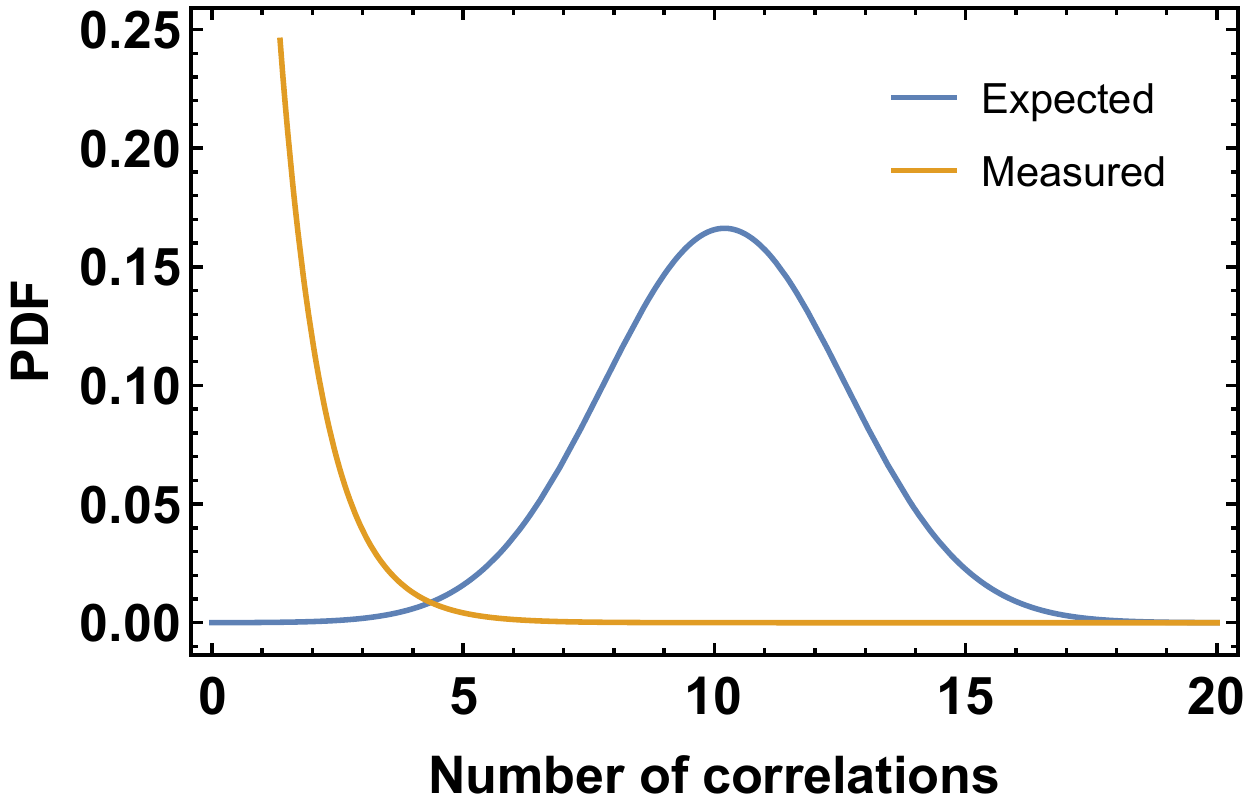}
\caption{\textit{The orange line is the PDF \cambia{of the observed number of correlations.}
In blue line we show the expected number of correlations, assuming that the BL Lacs are the main emitters of high-energy neutrinos,  
discussed after Eq.~\ref{sigexp}. }}
\label{pcdf}
\end{figure}

\andreab{In order to generalize the procedure, we scale the uncertainties given 
in Table~4 of \cite{muoni6} by a coefficient $k$, in the following manner:
\begin{equation}
\Delta(k)=\frac{k}{1.65}\times \Delta_{90\%}
,\end{equation}
where $\Delta_{90\%}$ are the uncertainties \andreac{on the declination and on the right ascension quoted in the table at 90\% C.L., $\Delta(k)$ are the new uncertainties of $\alpha$ and $\delta,$} and $k$ denotes the number of
(one dimensional) $\sigma$ of our interval. 
The value of the two-dimensional confidence level is given simply by the  square of the integral of the standard normal distribution between $-k$ and $k$, 
\begin{equation}
\mbox{C.L.}(\mbox{2 d.o.f},k) = \left[ \int_{-k}^{k} \frac{1}{\sqrt{2 \pi}} e^{-\frac{x^2}{2}} \ dx \right]^2
.\end{equation}
Some values of $k$, the corresponding confidence levels and the number of correlations observed within a certain C.L. are reported 
in Table~\ref{tab:kcl}. 
The number of correlations within a certain C.L. \cambiab{(see Table~\ref{tab:kcl})}\ is reasonably well fitted by a probability density function with an exponential shape.
Therefore the Probability Distribution Function (PDF) of the observed number of correlations is given by the derivative of
\begin{equation}\mathcal{L}_{obs}(n)  = 1- \exp \left(\frac{n}{n^*} \right)
\mbox{ with }n^*=0.90 \pm 0.05
.\end{equation} 
In Fig.~\ref{pcdf} we show how many correlations are present within a certain confidence level. 
The PDF is shown in Fig.~\ref{pcdf} by an orange line. The average number of correlations is 0.9 whereas the median value is 0.6. In the same figure the distribution of the expected number of correlations from theoretical considerations  is also illustrated (blue line).  }

\andreab{In order to compare two different distributions, that is, the expected theoretical correlations $\mathcal{L}_{th}(n)$ and the observed correlations $\mathcal{L}_{exp}(n)$, we use the same procedure described in \cite{gal1}, adopting the following formula to evaluate the \lq\lq distance\rq\rq \ between the distributions:
 \begin{equation}
 \mathcal{L}(\delta) = \int_0^\infty \ \mathcal{L}_{exp}(n) \times \mathcal{L}_{th}(n+\delta) \ dn
 .\end{equation}
 We found that $\mathcal{L}(\delta)$ is in good approximation and normally distributed, with a mean value of $\mu=9.4$ and a standard deviation of $\sigma=2.5$. The null value is excluded at $3.7 \sigma$ and this represents the difference between these two distributions. Taking into account the uncertainty in the fit of $\mathcal{L}_{th}(n)$ we found a similar result, since the two distributions are different with a significance of $3.7 \pm 0.1 \sigma$.}
Excluding the outliers, that is, the neutrino event numbers 6 and 14, the difference between the two distributions has a significance of 4.1$\sigma$. \andread{These considerations are true under the hypothesis that neutrinos are produced in the BL Lacs, associated to the $\gamma$-rays that we observe from those sources. We refer to \cite{Kalashev:2013vba} for an alternative scenario in which high-energy neutrinos are not produced directly in the source but are given by the interaction of cosmic rays produced in a generic AGN and the EBL.}

\andreac{Combining the information provided by the theoretical likelihood and the experimental likelihood, it is also possible to find the contribution of the BL Lacs to the IceCube neutrinos above 0.2 PeV. \corr{We can estimate the fraction $\xi$ of the events, that can be attributed to the BL Lacs, by means of the following likelihood function:}
\begin{equation}
\mathcal{L}_{frac}(\xi) \sim \int_0^\infty  \mathcal{L}_{exp}(\xi \times n) \mathcal{L}_{th}(n) \ dn
\label{eq10}
\end{equation}
In this way, we find that the fraction is $\xi=0.11$ at 67\% C.L. and $\xi=0.23$ at 90\% C.L. Considering the present data, the probability that the BL Lacs contribute more than 50\% to the through-going muons above 0.2 PeV is less than~1\%.    }

\begin{figure}[t]
\includegraphics[width=0.45\textwidth,angle=0]{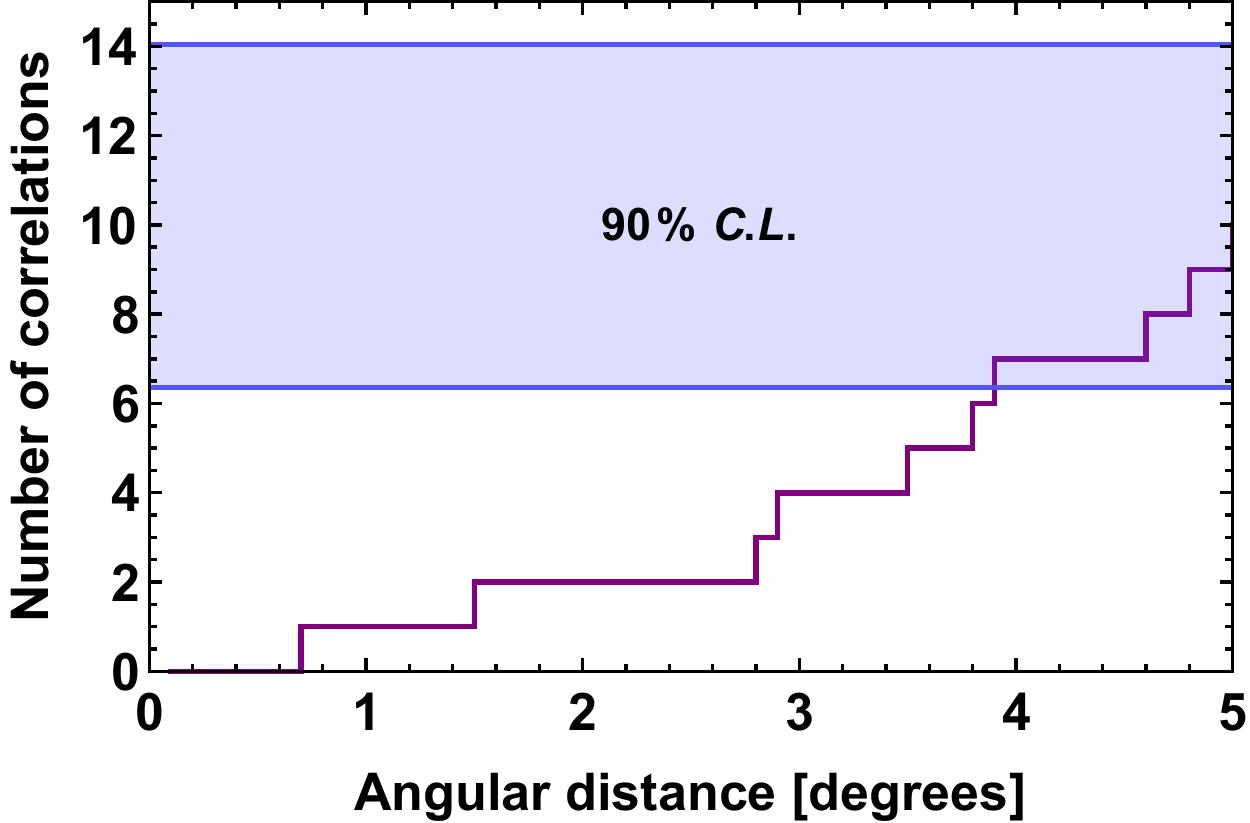}
\caption{\textit{
Number of correlations found as a function of the angular distance between the position of the neutrino events and the position of the BL Lac. Within 1 degree, only 1 correlation has been observed. The horizontal band shows the expectations given in Eq.~\ref{sigexp}. 
}}
\label{ncs}
\end{figure}

\subsection{Number of observed correlations estimated varying the  angular intervals} An alternative procedure is the following one. 
If we disregard the angular resolution quoted by the IceCube collaboration, and we vary it instead, 
it is interesting to ask 
how many events have a counterpart within a certain angular distance. Considering a neutrino event with coordinates $(\alpha_1,\delta_1)$ and a BL Lac with coordinates $(\alpha_2,\delta_2)$, the angular distance $d$ is given by the spherical distance, defined as follows:
\begin{equation}
d \equiv \sqrt{(\delta_2 - \delta_1)^2 + (\alpha_2- \alpha_1)^2 \cos(\delta_1)\cos(\delta_2)}
\label{eq:dist}
.\end{equation}
This expression can be obtained by  expanding  for $\delta_1\approx \delta_2$ and  $\alpha_1\approx \alpha_2$, 
the scalar product of the two unit vectors in the given directions, $\cos(d)\equiv (n_1,n_2)$. Therefore, $d$ subtends the region of solid angle 
\begin{equation}
\Omega=\pi d^2
\label{eq:angres2}
,\end{equation}
and it is directly comparable with the average angular resolution $\theta$ defined in Eq.~(\ref{defello}).

Using this procedure we find that 
only one event has a counterpart at distance \andreac{$d \leq 1^\circ$}; it is the through-going event number 23 of Table~4 of \cite{muoni6} that can be correlated with the BL Lac \textit{J0211.2+1050}. 
We have tested that within \cambiab{$d \leq 2^\circ$} there are two possible counterparts.
Finally, for completeness, we report in  Tables~\ref{default} the 
4 events\footnote{We note that the two correlations between the through-going muon number 6 and the BL Lacs are not present in this table, since the angular distance between them is larger than $3^\circ$, using the best fit coordinates of the neutrino directions. However, the very large uncertainties would not permit us to rule out the correlations between them.} that show at least one correlation with a BL Lac \andreac{within $\sim 3^\circ$}. 

\andreab{\andreac{In Fig.~\ref{ncs} we present} a generalization of this analysis}. Assuming that the BL Lacs are the counterpart of the IceCube neutrino events, we wonder how much  the angular resolution of the tracks should be worsened, in order to observe a certain number of correlations.

\andreac{If we want to recover the agreement with the expectation at 90\% C.L., we need an average angular distance of $d \simeq 4^\circ$, namely, $\Omega=50$ square degrees (eq.\ref{eq:angres2}). This value is very different from the angular resolution declared for the tracks, typically close to $1^\circ$ and more precisely equal to $\bar{\theta}=1.2^\circ \pm 0.9^\circ$ for the present through-going muons dataset, as discussed after Eq.~(\ref{defello}). Therefore the required angular resolution of $4^\circ$ is in tension with the declared angular resolution with a significance of $3\sigma$ 
and could only be justified invoking a very large value of the systematic uncertainty.   }




\begin{table}[b]
\caption{\textit{Four events are reported here that show correlation with BL Lacs within $3^\circ$. \cambiab{The angular distance is defined in Eq.~(\ref{eq:dist}).}}}
\begin{center}
\begin{tabular}{ccc}
\hline
 ID $\mu$ event & BL Lac & \cambiab{$d$} \\
\hline
23 & J0211.2+1050 & $< 1^\circ$ \\
1 & J0152.8+0146  & $< 2^\circ$ \\
2 & J1942.8+1033 & $<3^\circ$ \\
18 & J2153.1-0041 & $<3^\circ$ \\
\hline
\end{tabular}
\end{center}
\label{default}
\end{table}%

\omitt{
\subsection{Discussion}
\label{discbl}
The observation of only 1 neutrino associated to a BL Lac does not fit  well with the expectations given in Eq.~\ref{sigexp}. If we consider the probability to observe 1 event or less with  Poisson statistics, $p_1(\mu)=e^{-\mu}(1+\mu)$ we get $p_1(10.2)=4\times 10^{-4}$ which amounts to 3.5$\sigma$. Accounting for the uncertainties in the expectations by a Gaussian distribution $G(\mu'-\mu,\sigma)$ with $\sigma=2.4$,  
\begin{equation}
p_1(\mu,\sigma)= \int_0^\infty p_1(\mu') \times G(\mu'-\mu,\sigma) \ d\mu'
\end{equation}
we find a probability smaller than 1\%, i.e.,
\begin{equation}P(\leq 1 \ | \ G(N_s^c,\Delta N_s^c)) = 0.0036 \ ;\end{equation}
this scenario is disfavored at $2.9 \sigma$.
Finally, considering the 90\% C.L.\ lower limit in the expectations, in order to be conservative, we find a $2.5 \sigma$ exclusion of the hypothesis. 

In the left panel of Fig.~\ref{ncs} we generalize this calculation, reporting the number of $\sigma$ of exclusion as a function of the observed number of events that show correlation with a source. 
The dashed red line shows the present situation, with only one event matching the assumption. 
\omitt{
The probability to observe 1 event or less is given by the Poissonian statistics, $P(\mu)=e^{-\mu}(1+\mu)$, weighted by the distribution $G(\mu'-\mu,\sigma)$, as follows:
\begin{equation}
P(\leq 1 \ | \ G(\mu'-\mu,\sigma))= \int_0^\infty P_1(\mu') \times G(\mu'-\mu,\sigma) \ d\mu'
\end{equation}

Using the values found in the previous paragraph, $\mu=N_s^c$ and $\Delta \mu=\Delta N_s^c$, we found a probability smaller than 1\%, i.e.
\end{equation}P(\leq 1 \ | \ G(N_s^c,\Delta N_s^c)) = 0.0036 \ ;\end{equation}
this scenario is disfavored at $2.9 \sigma$.

\andrea{Even considering the lower limit at 90\% C.L.\ of Eq.~(\ref{sigexp}) instead of the distribution, in order to be more conservative, we found a $2.5 \sigma$ exclusion}
In the left panel of Fig.\ref{ncs} we generalize this calculation, reporting the number of $\sigma$ of exclusion as a function of the observed number of events that show correlation with a source. We use both the lower limit at 90\% C.L. of our expectation (orange line) and the central value (blue line). The dashed red line shows the present situation., with only one observation. 
}

For completeness, we have also tested the tracks contained into the last HESE dataset  collected in the first 4 years given in \cite{hese}, finding no counterparts in the $\gamma$-rays sky within $2^\circ$. Anyway, we prefer to omit this data set from our analysis, since the energy threshold of HESE is 30 TeV and at this energy the contamination from atmospheric background is much more relevant. In this manner, our conclusions can be considered as conservative.  
}

\begin{figure}[t]
\centering
\includegraphics[width=0.45\textwidth,angle=0]{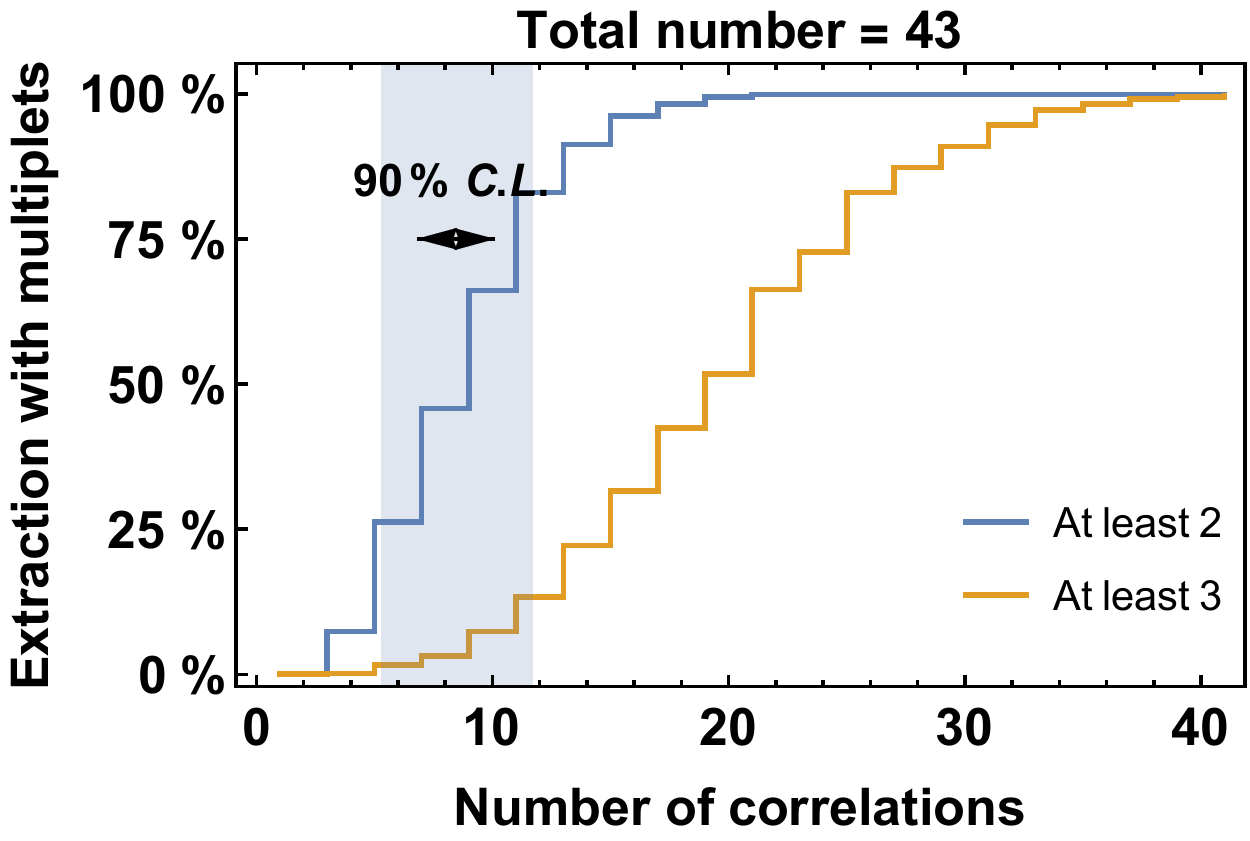}
\caption{\textit{Probability of observing one multiplet as a function of the number of correlations. 
We consider the 43 
BL Lacs 
contained in the 2FHL Fermi-LAT catalog in the region $0^\circ \leq \delta \leq 30^\circ$. \andrea{The luminosities of BL Lacs are taken into account in the calculation of the expectations.} 
The expected number of correlations,  assuming that the BL Lacs  are the main emitters of high-energy neutrinos (Eq.~\ref{cartiz}), 
is indicated by the vertical band.}}
\label{mult}
\end{figure}

\caga{
\section{Multiplets}
\label{multi}
Here, we discuss a more refined test of the hypothesis  that BL Lacs are the main emitters of high-energy neutrinos: 
We evaluate the probability of observing at least one multiplet, that is, two or more events from the same source. \andrea{In this analysis we refer to identified BL Lacs contained in the second Fermi-Lat catalog (\cite{2fhl}).}

As is clear from Fig.~\ref{map},  
most through-going muons come from the region $0^\circ \leq \delta \leq 30^\circ$. 
This is expected for the high-energy neutrino signal, 
due to the absorption in the Earth; this conclusion is 
consistent with the discussion given in \cite{muoni6}.
Therefore we limit our analysis to the cleaner subset of 
BL Lacs that are contained in this part of the Northern sky. 
Repeating the same calculation of Sec.~\ref{liksig}, for the subset of 23 events included in the interval of declination $0^\circ \leq \delta \leq 30^\circ$, we find that the number of events that can be attributed to cosmic neutrinos is equal to,
$
N_{\mbox{\tiny signal}}(0^\circ \leq \delta \leq 30^\circ)=16.9 \pm 2.1
$
at 1$\sigma$. 
Using Eq.~(\ref{evcorr}) we find that the number of signal events, expected to show a correlation due to their common origin, is equal to
\begin{equation} \label{cartiz}
N_{\mbox{\tiny corr}}(0^\circ \leq \delta \leq 30^\circ)=8.5 \pm 2.0
,\end{equation} 
at a C.L. of 1$\sigma$ , which becomes $8.5 \pm 3.2$ at 90\% C.L.


In this angular region, 43 BL Lacs  are present in the Fermi-LAT catalog (\cite{2fhl}).
Assuming that the probability of emitting a neutrino is proportional to the luminosity of the sources, as argued in \cite{TavecchioRighi}, \andrea{we define a weight $w_i$ as follows}:
\begin{equation}
w_i=\frac{L_i}{L_{tot}}
,\end{equation} 
where $L_i$ and $L_{tot}$, taken from \cite{2fhl}, are the luminosity of a single source and the sum of luminosity of the 43 sources, respectively.
\andrea{In order to evaluate the expected number of multiplets, as a function of the number of individual correlations, we perform several cycles of random extractions of 43 numbers, each one with a probability of extraction equal to $w_i$. Each one of these extractions can be thought of as a simulated experiment.  
Then, we count the frequency of occurrence of a multiplet with at least two or three events from the same source.\footnote{\andrea{In order to test the stability of our procedure, we also repeated the same calculation using a uniform extraction of the numbers, finding a very similar result. }}
}

The outcome is illustrated in Fig.~\ref{mult}.
With the expected number of individual correlations, 
shown as a vertical band in the figure, 
the probability of non observation of a multiplet from a single source is between 25-75\%. Thus, the fact that we do not observe any multiplet is not yet an issue. The lack of such an 
observation would become significant if the expected number of events was instead of the order of about 15-20.
Therefore, this test will become important when the current 6-year statistics are doubled, or when we  have 2-3 years of data from IceCube-Gen2,  due to the bigger effective area of the incoming detector described in \cite{ice2}.}

\andread{This analysis strongly depends on the class of objects considered and the absence of multiplets is not (yet) a problem assuming that high-energy cosmic neutrinos, above 0.2 PeV, are fully produced by BL Lacs. We refer to \cite{murase2} for a scenario in which, on the contrary, the absence of multiplets represents an issue even with the present IceCube data. }

\section{Summary and discussion}
\label{conc}
\andreab{The BL Lacs are potentially good candidates for emitters of high-energy neutrinos. The existing upper bound on neutrino flux emitted  from the resolved blazars, obtained by the IceCube collaboration (\cite{icecube1}), is not incompatible with the assumption that the BL Lac\corr{s} account for the full high-energy part of the spectrum, measured by means  
of the through-going muons above 0.2 PeV, described in \cite{muoni6}. }

We have performed a \andreac{refined} test of this hypothesis. 
Among the through-going muon dataset, approximately two thirds of the events should be attributed to the astrophysical neutrinos. Assuming that these are produced in BL Lacs, we find that half of them, that is, approximately one third of the complete dataset, should be correlated with a BL Lac of the second Fermi-LAT catalog (2FHL).   This leads us to expect some ten correlations with an uncertainty of about 25\%. \andreab{However we found that \cambia{only the event number 6 has a BL Lac counterpart at 81\% C.L. Even enlarging the 
search window, in order to reach 99\% C.L., we find only three possible correlations.  Generalizing,}
the average number of tracks that can be correlated with known BL Lacs is well below expectations.

This represents a serious issue for the hypothesis that the BL Lacs are the \andreac{main} sources \andreab{of neutrinos above 0.2 PeV.} 
Assuming that the systematic uncertainty (quoted in \cite{muoni6} only for the most energetic event) is smaller than the statistical uncertainty,} this hypothesis 
turns out to be disfavored at about 3.7$\sigma$. \andreab{Moreover, the direction of event number 6, that shows a correlation within 81\% C.L.,\ is not well reconstructed, since the uncertainty on its direction is of the order of \andreac{$10^\circ$}. Therefore the correlation with the BL Lac could simply be a coincidence due to poor angular resolution. } If the result discussed above is not due to a statistical fluctuation and there are no unaccounted systematic effects, the most plausible inference is that 
the BL Lacs are not the main emitters \andreab{of the high-energy neutrinos above 0.2 PeV}. 

\corr{In view of the significance of this conclusion, it is important to discuss possible ways-out from the strong bound on the possible high-energy neutrino emission that we have obtained, that implies  the conclusion that the observed neutrinos with energy above 200 TeV receive only a minor contribution from BL Lacs.}
We have shown that this hypothesis could be reconciled with the observations if the angular resolution of the tracks in IceCube was not as good as is quoted. 
The angular resolution of track-like events should be of $\sim 4^\circ$  to reconcile the expectations with the observations.
The required departure is quite large and this  
makes this interpretation less plausible. 
For this reason, we are inclined to believe that 
this result argues in favor of the hypothesis  that the high-energy neutrinos come from a population of faint neutrino emitters, meaning that most of the  neutrino-emitting sources are not resolved with the present instrument; 
from a truly diffuse source (e.g., the halo of our Galaxy as argued in \cite{alone}); or from a type of source where $\gamma$-rays are heavily reprocessed and the radiation shifted at much lower energies.
\att{We note that it is also not possible to exclude the theory that BL Lacs themselves, as identified by $\gamma$-ray astronomy,
 are the sources of the highest-energy neutrinos seen by IceCube, if the neutrinos are emitted in a significantly wider angular range;
this would imply a drastic departure from the spine-layer model of  \cite{tavecchio14b,tavecchio14a,TavecchioRighi}.} 
\andreab{Finally, we remark that we cannot even exclude the possibility that BL Lacs are good emitters of neutrinos below 0.2 PeV, since this energy region has not been investigated in the present work.}


In this paper, we have also tested the probability of observing at least two events from the same source, assuming that BL Lacs are the high-energy neutrino emitters. Our result is that the non observation of multiplets does not represent an issue for the BL Lac hypothesis nowadays, but will become crucial in the  future, especially after IceCube-Gen2 begins collecting data.

To conclude, some important remarks are in order: 
First, we highlight that our result \andreab{complements and strengthens the recent upper bound on the blazar emission derived by IceCube in \cite{icecube1}. 
\cambia{Second,  we have argued that a correct 
understanding of the true 
spectrum of the extragalactic neutrinos is of  utmost importance for multi-messenger astronomy.} 
Third, we note that multi-messenger adopted in this work has great scientific 
potential and will allow us to proceed in the study of the origin of astrophysical neutrinos observed by IceCube, also 
for other types of astrophysical sources. }




\omitt{
\att{\begin{itemize}
\item i bllac vanno bene a livello spettrale (saturano);
\item assumendo che l'errore formale riportato in tabella e che l'errore sistematico sia piccolo arriviamo alla conclusione che i bllac non correlano spazialmente con gli eventi di IceCube;
\item se errore è molto più grande di quello riportato in tabella è possibile che bllac sia anche la sorgente di più alta energie
\item i multipletti non sono ancora un buon sistema per discriminare l'ipotesi bllac. Al momento non pregiudicano analisi
\item non c'e' contraddizione con la nostra analisi e quella di icecube. e' importante precisione angolare ma e' anche importante chiarire l'interpretazione degli eventi di bassa energia (comp. galattica, background ecc) . Serve per fare multimessenger in modo corretto
\end{itemize}} }

\vspace{0.3 cm}

\paragraph*{Acknowledgements}
We thank E.~Resconi, S.~Celli, M.~Spurio, V.~Kudryavtsev, A.~Esmaili, M.~Tavani, F.~Tavecchio and C.~Righi for precious discussions.

\bibliographystyle{apsrev4-1}
\bibliography{bibliografia}



\end{document}